\documentclass[aps,twocolumn,pra,twoside,amssymb,amsmath]{revtex4}
\usepackage{amssymb}
\usepackage{graphicx}
\usepackage{amsmath}
\usepackage{colordvi}
\usepackage{bbm}
\usepackage{verbatim}
\usepackage{float}
\usepackage{dcolumn}
\usepackage[colorlinks,linkcolor=blue,anchorcolor=blue,citecolor=blue,urlcolor=blue]{hyperref}
\usepackage{mathrsfs}

\begin{document}
\title{High-purity fluorescence photon bundle emission from two separate emitters}
\author{Zhicai Chen$^{1}$, Jun Xu$^{1}$, Deyi Kong$^{2}$,  Xiangming Hu$^{1,}$\footnote{xmhu@ccnu.edu.cn} and Fei Wang$^{2,}$\footnote{feiwang@hbut.edu.cn}}
\affiliation{
  $^{1}$\mbox{College of Physical Science and Technology, Central China Normal University, Wuhan 430079, China} \\
  $^{2}$\mbox{School of Science, Hubei University of Technology, Wuhan 430068, China}}
\begin{abstract}
We propose a scheme for generating high-purity fluorescence two-photon bundles from two spatially separate two-level emitters, whose correlated excitation is mediated by a strongly driven auxiliary emitter. Unlike bosonic-mode-based approaches, the finite excitation space of the two target emitters intrinsically excludes higher-excitation manifolds, and thus eliminates impurity photons associated with undesired higher-excitation states. We further analytically characterize the residual population of off-resonant single-excitation states and show that suppressing the corresponding leakage improves photon bundle purity. Our scheme offers a feasible pathway for constructing high-purity quantum light sources with promising applications in quantum information processing.
\end{abstract}

\maketitle
%==============================================================================================================
\section{introduction}\label{sec1}

The generation and manipulation of photon bundle emission is an important task in multiphoton physics because of its potential applications in quantum information processing \cite{Bennett2000,Couteau2023,Kimble2008,Ladd2010}, quantum metrology \cite{Dowling2008,Joo2011,Qin2023,Giovannetti2011,Paris2009}, multiplexed quantum communication \cite{MeyerScott2022,Gisin2007,Sangouard2011}, entanglement generation \cite{Llewellyn2020}, and ultrasensitive biosensing \cite{Horton2013,Li2023}. Up to now, various systems have been proposed to realize bundle emission, including cavity-QED systems \cite{munoz2014emitters,Munoz2018,Gou2022,Gou2024NJP,Jiang2023,Tang2023,Liu2023}, semiconductor quantum dots \cite{Munoz2015,Cosacchi2022}, waveguide-QED systems \cite{GonzalezTudela2017,Xing2024}, superconducting circuit platforms \cite{Ma2021,Ma2022,Zou2023}, phononic and magnonic systems \cite{Bin2020,Zou2022,Deng2021,Yuan2023,Liu2025OptLett}, and hybrid magnon--photon or waveguide-magnonics platforms \cite{Gou2024PRR,Bin2024LPR,Wang2025OL,Zhao2025PRA}. Related cavity-based schemes have also been developed for deterministic and continuous generation of correlated photon pairs \cite{Chang2016}.
A representative mechanism in cavity-QED systems relies on an effective multiphoton transition that coherently couples the ground state to a target higher-excitation state, giving rise to a super-Rabi oscillation \cite{munoz2014emitters,Munoz2018}. The photons are then emitted sequentially within each bundle, while successive bundles exhibit antibunching. The performance of such bundle sources is largely determined by both the emission brightness and the bundle purity. Increasing the driving strength can enhance the excitation of the target multiphoton state, but excessively strong driving may also populate unwanted higher-excitation states and induce re-excitation processes, thereby degrading the bundle purity. Therefore, suppressing unwanted excitation channels while maintaining efficient bundle emission remains an important challenge.

Near-unity two-photon bundle purity can be achieved in cavity-QED schemes under suitably optimized parameters \cite{munoz2014emitters}. In more general parameter regimes, however, photons emitted from unwanted Fock-state components may contaminate the target $N$-photon bundles and reduce their purity. Several strategies have therefore been developed to enhance the desired bundle-emission channel while suppressing unwanted emission processes. For example, parametric amplification can enhance the effective light--matter coupling and thereby increase the excitation of target multiphoton states and the bundle emission rate \cite{Chen2025PRA}. Under appropriate conditions, parity protection can favor even-order $2n$-photon bundle emission while suppressing odd-photon emission, thereby substantially improving the purity of the even-photon bundles \cite{Bin2021,Xiong2025}. Dark-state engineering assisted by an ac Stark shift provides another route to high-quality multiphoton bundle emission by establishing super-Rabi oscillations involving $N$-excitation dark states \cite{Gou2024NJP}. In a different approach, a chiral non-Hermitian interaction in a cascaded-cavity system can prolong the lifetime of the target state and suppress re-excitation, thereby improving both the bundle purity and emission number \cite{Chen2026Chiral}. Nevertheless, these approaches are still based on bosonic modes with an unbounded Fock-state ladder, so unwanted higher-excitation components must be dynamically suppressed rather than intrinsically excluded.

To overcome this limitation, we propose a different strategy for generating photon bundles using three coupled two-level emitters. Two spatially separate target emitters are individually coupled to a strongly driven auxiliary emitter, while no direct coupling exists between the target emitters. In the dressed-state picture, the auxiliary emitter mediates an effective correlated excitation of the two target emitters, which subsequently emit a fluorescence two-photon bundle. The key advantage of the present scheme is that each target emitter can accommodate at most one excitation, so that the joint target-emitter subspace intrinsically excludes higher-excitation manifolds, thereby eliminating impurity photons associated with higher-excitation states. The remaining imperfection originates mainly from the residual population of the off-resonant single-excitation states, whose decay gives rise to undesired single-photon emission. We analytically show that this leakage is suppressed as $g^2/\Omega^2$ in the strong-driving regime, allowing the bundle purity to approach $100\%$ as the driving strength of the auxiliary emitter increases. At the same time, the effective correlated-excitation strength decreases as $|g_{\rm eff}|\sim g^2/\Omega$, leading to a trade-off between bundle purity and emission rate. Finally, we discuss a possible implementation of the present scheme in superconducting quantum circuits.

This paper is organized as follows. In Sec.~\ref{sec2}, we introduce the three-emitter model and present the system Hamiltonian in the dressed-state picture. 
Sec.~\ref{sec3} investigates the fluorescence two-photon bundle emission, characterizing correlations and visualizing dynamics via quantum trajectories. 
Sec.~\ref{sec4} focuses on suppressing off-resonant single-excitation states, quantifying their impact on bundle purity, and exploring the effect of driving strength on single-photon leakage. Finally, Sec.~\ref{sec5} summarizes our conclusions.
%==============================================================================================================
\section{Model and equations}
\label{sec2}

\begin{figure}[t]
  \centering
  \includegraphics[scale=0.6]{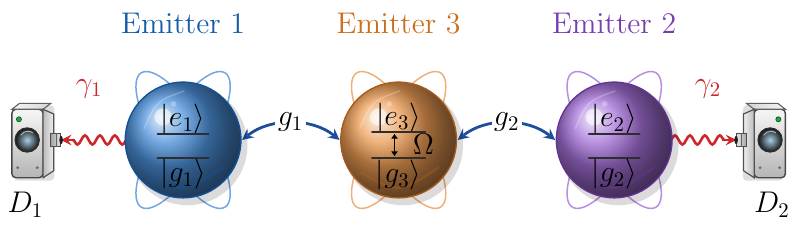}
  \caption{
  Schematic of the system consisting of three artificial two-level emitters. 
  Emitters~1 and~2 are the target emitters, whose fluorescence photons form a two-photon bundle. 
  Emitter~3 is an auxiliary emitter coherently driven by a classical field and coupled to both target emitters, while there is no direct coupling between emitters~1 and~2. 
  The fluorescence photons emitted by emitters~1 and~2 are collected by detectors $D_1$ and $D_2$, respectively, for photon-correlation measurements.
  \label{Fig1}
  }
\end{figure}

The system under consideration is schematically shown in Fig.~\ref{Fig1}. We consider three artificial two-level emitters that can be implemented in superconducting circuits. Emitters~1 and~2 serve as the target fluorescence emitters, while emitter~3 is an auxiliary emitter coherently driven by an external classical field. The auxiliary emitter is coupled individually to the two target emitters, whereas no direct coupling is introduced between emitters~1 and~2. Such coherent exchange couplings can be implemented through capacitive or inductive interactions, or effectively engineered using tunable couplers in superconducting circuits \cite{Niskanen2007,Bialczak2011,Zhang2024}. A closely related three-qubit coupling configuration has also been experimentally realized with three frequency-tunable transmon qubits, where one qubit is effectively coupled to each of the other two through tunable couplers \cite{Li2024ThreeQubit}.

In the frame rotating at the driving frequency $\omega_d$, the system Hamiltonian ($\hbar=1$) reads
\begin{align}
H &= \sum_{j=1}^3 \Delta_j \sigma_j^\dagger\sigma_j 
   + \sum_{j=1}^2 g_j (\sigma_j^\dagger\sigma_3 + \sigma_j \sigma_3^\dagger) \nonumber \\
  &\quad + \Omega (\sigma_3^\dagger + \sigma_3).
\label{Eq1}
\end{align}
where $\sigma_j = |g_j\rangle\langle e_j|$ and $\sigma_j^\dagger = |e_j\rangle\langle g_j|$ are the lowering and raising operators of emitter $j$, respectively. 
Here $\Delta_j = \omega_j - \omega_d$ is the detuning of emitter $j$ from the driving field, $g_1$ ($g_2$) denotes the coupling strength between emitter~1 (2) and emitter~3, and $\Omega$ is the Rabi frequency of the coherent drive on emitter ~3.

The dissipative dynamics of the system is governed by the Lindblad master equation
\begin{equation}
\label{master}
\frac{d\rho}{dt} = -i[H,\rho] + \sum_{j=1}^{3} \frac{\gamma_j}{2} 
\bigl(2\sigma_j\rho\sigma_j^\dagger - \sigma_j^\dagger\sigma_j\rho - \rho\sigma_j^\dagger\sigma_j\bigr),
\end{equation}
where $\gamma_j$ is the decay rate of emitter $j$. In particular, the fluorescence output is collected from emitters~1 and~2, whose emitted photons characterize the two-photon bundle emission.

To generate fluorescence photon bundle emission, we operate in the Mollow regime of strong driving $(\Omega\gg g_1,g_2)$. Therefore, it is convenient to elucidate the underlying mechanism in the dressed-state picture of the driven emitter ~3. The Hamiltonian of the strongly driven emitter ~3 can be diagonalized as $\tilde{H}_3=\lambda_+\sigma_{++}+\lambda_-\sigma_{--}$, where $\sigma_{\alpha\beta}=|\alpha\rangle\langle\beta|$ $(\alpha,\beta=\pm)$, $\lambda_{\pm}=\frac{1}{2}(\Delta_3\pm\Omega_R)$, and $\Omega_R=\sqrt{\Delta_3^2+4\Omega^2}$. Then, the dressed states can be expressed in terms of the bare states as follows:
\begin{eqnarray}
\label{dressed_states}
\nonumber
|+\rangle &=& \sin\theta |g_3\rangle+\cos\theta |e_3\rangle,\\
|-\rangle &=& \cos\theta |g_3\rangle-\sin\theta |e_3\rangle,
\end{eqnarray}
where
$\sin\theta=\sqrt{\frac{1}{2}-\frac{d}{2\sqrt{1+d^2}}}$ and
$\cos\theta=\sqrt{\frac{1}{2}+\frac{d}{2\sqrt{1+d^2}}}$ with
$d=\Delta_3/(2\Omega)$.
Based on the dressed-state picture, the resonant transition channels and corresponding resonance conditions are obtained as follows, with the detailed derivation given in Appendix~\ref{app:effective_hamiltonian}:
\begin{align}
\nonumber
|g_1 g_2, +\rangle &\leftrightarrow |e_1 e_2, -\rangle, 
\quad \Delta_1 + \Delta_2 \approx \Omega_R, \\
|g_1 g_2, -\rangle &\leftrightarrow |e_1 e_2, +\rangle, 
\quad \Delta_1 + \Delta_2 \approx -\Omega_R.
\label{resonance_channels}
\end{align}
In these two channels, the simultaneous excitation of emitters~1 and~2 is mediated by the opposite dressed-state transitions of emitter~3, i.e., $|+\rangle \to |-\rangle$ and $|-\rangle \to |+\rangle$.

\begin{figure}[t]
  \centering
	\includegraphics[scale=0.4]{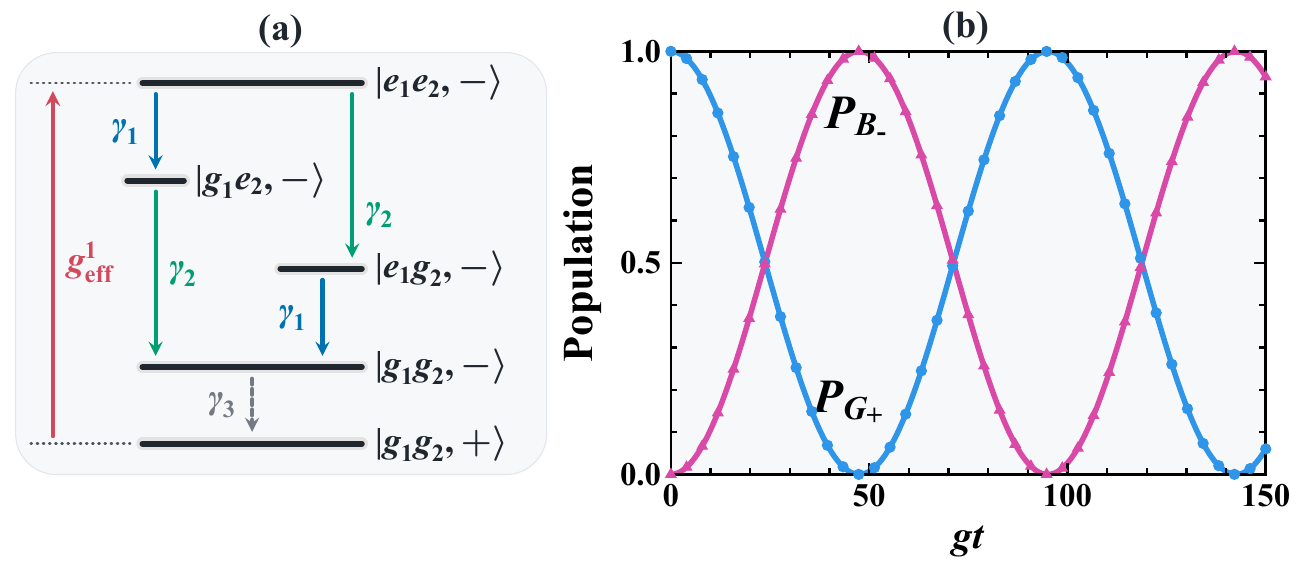}
\caption{ (a) Mechanism schematic for fluorescence photon bundle emission.
(b) Time evolution of the populations $P_{G_+}$ and $P_{B_-}$ under the resonance condition $\Delta_1+\Delta_2\approx \Omega_R$ in the absence of dissipation. 
The solid lines represent the numerical results obtained from the full Hamiltonian in Eq.~\eqref{Eq1}, while the triangular and circular symbols denote the analytical results based on the effective Hamiltonian in Eq.~\eqref{eq:Heff_main}. 
The initial state is $|G_+\rangle$. 
The parameters are $g_1=g_2=g$, $\Delta_1=\Delta_2=30g$, $\Delta_3=0$, and $\Omega=30g$.
\label{Fig2}}
\end{figure} 

\begin{figure*}[t]
  \centering
  \includegraphics[width=\textwidth]{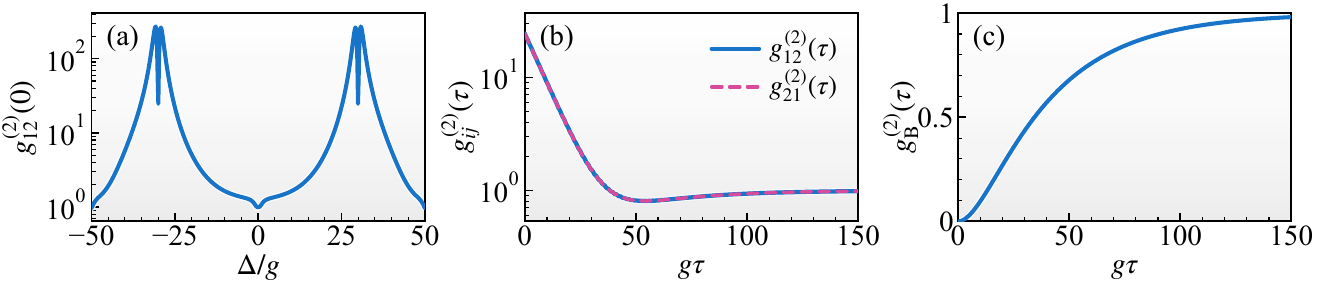}
 \caption{
Second-order correlation functions of the fluorescence photons emitted from the two target emitters. 
(a) Equal-time cross-correlation function $g_{12}^{(2)}(0)$ as a function of the normalized detuning $\Delta/g$. 
(b) Delayed cross-correlation functions $g_{12}^{(2)}(\tau)$ and $g_{21}^{(2)}(\tau)$ at the resonant point $\Delta=30g$. 
(c) Delayed bundle correlation function $g_{\rm B}^{(2)}(\tau)$ at the resonant point $\Delta=30g$. Other parameters are $\Omega=30g$, $\gamma=0.1g$ and $\gamma_3=0.1\gamma$.
} \label{Fig3}
\end{figure*}

In the following, we focus on the first resonant channel, while the second channel can be treated in the same manner. 
For clarity, we introduce the shorthand notations 
$|G_\pm\rangle\equiv |g_1g_2,\pm\rangle$ for the joint ground states, 
$|X_{1\pm}\rangle\equiv |e_1g_2,\pm\rangle$ and 
$|X_{2\pm}\rangle\equiv |g_1e_2,\pm\rangle$ for the single-excitation states, and 
$|B_\pm\rangle\equiv |e_1e_2,\pm\rangle$ for the two-excitation states.
By perturbatively eliminating the off-resonant single-excitation states, the effective Hamiltonian for this channel is obtained as
\begin{align}
H_{\rm eff}^{1}
&= g_{\rm eff}^{1} \bigl( \sigma_1^\dagger \sigma_2^\dagger \sigma_{-+} + \sigma_1 \sigma_2 \sigma_{+-} \bigr) \nonumber \\
&= g_{\rm eff}^{1} \bigl( |B_-\rangle\langle G_+| + |G_+\rangle\langle B_-| \bigr),
\label{eq:Heff_main}
\end{align}
with $g_{\rm eff}^{1}=-g_1g_2sc^3\left[\frac{1}{\Omega_R-\Delta_1}+\frac{1}{\Delta_1}+\frac{1}{\Omega_R-\Delta_2}+\frac{1}{\Delta_2}\right]$, and $c=\cos{\theta},\, s=\sin{\theta}$.
Under the  resonance condition $\Delta_1+\Delta_2\approx\Omega_R$, the effective Hamiltonian coherently couples $|G_+\rangle$ and $|B_-\rangle$, corresponding to the transition 
$|g_1g_2,+\rangle\leftrightarrow |e_1e_2,-\rangle$. 
In this process, emitters~1 and~2 are excited simultaneously, while emitter~3 undergoes the dressed state transition $|+\rangle\rightarrow|-\rangle$. 
The corresponding fluorescence photon bundle emission process is schematically shown in Fig.~\ref{Fig2}(a). 
After the system is driven to the two-excitation state $|B_-\rangle$, as indicated by the red solid arrow, emitters~1 and~2 release the target two-photon bundle via spontaneous emission, indicated by the blue and green solid arrows, respectively. These two photons are strongly correlated, as they are emitted one after the other from the same two-excitation state. The emission order of the two target emitters is random. The subsequent emission from emitter~3, indicated by the gray dotted arrow, returns the system to its initial state and separates the present bundle from the next bundle emission cycle.

To verify that this effective transition path indeed captures the dominant dynamics, we examine the population evolution in the absence of dissipation. 
As shown in Fig.~\ref{Fig2}(b), starting from $|G_+\rangle$, the populations of $|G_+\rangle$ and $|B_-\rangle$ undergo a nearly complete Rabi oscillation, confirming that the dynamics is restricted to these two states, while transitions to other states are effectively suppressed. 
The analytical results from Eq.~\eqref{eq:Heff_main} agree well with the numerical simulation using Eq.~\eqref{Eq1}, confirming the accuracy of the perturbative treatment.

An important advantage of the present two-emitter system is that the relevant Hilbert space is intrinsically finite. In contrast to bosonic modes, where higher number states may be populated, the two target emitters cannot support excitations beyond $|e_1e_2\rangle$. The finite-level structure therefore intrinsically excludes higher-excitation manifolds and the associated impurity photons. Ideally, the system dynamics is confined to the subspace $\{|G_+\rangle,|B_-\rangle\}$. In practice, however, the effective transition is mediated by the off-resonant single-excitation states $|X_{1\pm}\rangle$ and $|X_{2\pm}\rangle$, which can acquire small residual populations during the actual dynamics. Such intermediate-state leakage arises naturally because the effective two-emitter transition proceeds through higher-order virtual pathways. The radiative decay of these residual single-excitation states produces undesired single photons and therefore constitutes the dominant intrinsic contamination channel of the two-photon bundle emission. Having clarified the coherent two-emitter excitation mechanism, we next include emitter dissipation and investigate the full fluorescence emission dynamics.

\section{fluorescence photon bundle emission} \label{sec3}

With the two-photon bundle emission mechanism established, we now incorporate system dissipation and investigate the full fluorescence emission dynamics. 
To characterize the photon statistics of the target emitters, we introduce the delayed second-order correlation function
\begin{equation}
g_{ij}^{(2)}(\tau)
=
\frac{
\langle
\sigma_i^\dagger(0)\sigma_j^\dagger(\tau)
\sigma_j(\tau)\sigma_i(0)
\rangle
}{
\langle\sigma_i^\dagger(0)\sigma_i(0)\rangle
\langle\sigma_j^\dagger(\tau)\sigma_j(\tau)\rangle
},
\qquad i,j=1,2 .
\label{eq:g2_ij}
\end{equation}
For $i=j$, Eq.~\eqref{eq:g2_ij} reduces to the standard autocorrelation function of an individual target emitter. Since each target emitter is a two-level system satisfying $\sigma_i^2=0$, one has $g_{ii}^{(2)}(0)=0$, reflecting perfect antibunching of the fluorescence from each individual emitter.
For $i \neq j$, $g_{ij}^{(2)}(\tau)$ measures the temporal correlation between photons from emitters~$i$ and~$j$. When $\tau = 0$, $g_{ij}^{(2)}(0)$ denotes the equal-time cross-correlation. 
As defined, the bunching effect occurs when  $g_{ij}^{(2)}(0) > 1$.

The correlation between different photon bundles can be further characterized by the delayed bundle correlation function. 
By defining the joint emission operator $S=\sigma_1\sigma_2$, we write
\begin{equation}
g_{\rm B}^{(2)}(\tau)
=
\frac{
\langle
S^\dagger(0)S^\dagger(\tau)
S(\tau)S(0)
\rangle
}{
\langle S^\dagger(0)S(0)\rangle
\langle S^\dagger(\tau)S(\tau)\rangle
}.
\label{eq:g2_bundle}
\end{equation}

Here, $\tau$ denotes the time-delay. Since $S^2=0$ for two-level emitters, the literal zero-delay value $g_{\rm B}^{(2)}(0)=0$ as a consequence of the finite-level structure. 
We therefore use the delayed behavior of $g_{\rm B}^{(2)}(\tau)$ to characterize the antibunching between successive photon bundles.

For simplicity, we consider symmetric parameters 
$\Delta_1=\Delta_2=\Delta$, $\Delta_3=0$, $g_1=g_2=g$, and $\gamma_1=\gamma_2=\gamma$. 
The decay rates are set to $\gamma = 0.1g$ and $\gamma_3 = 0.1\gamma$, a choice typical in bundle emission studies. Figure ~\ref{Fig3}(a) shows the equal-time cross-correlation function $g_{12}^{(2)}(0)$ as a function of the normalized detuning $\Delta/g$. 
Two pronounced resonant features appear around $\Delta\simeq \pm 30g$, corresponding to the two resonant transition channels discussed above. 
In both regions $g_{12}^{(2)}(0) > 1$, indicating strong correlations between the fluorescence photons emitted from emitters~1 and~2. 
Focusing on the resonance condition $\Delta=30g$, we calculate the delayed cross-correlation functions in Fig.~\ref{Fig3}(b). 
Under the symmetric parameter conditions, the exchange symmetry of emitters~1 and~2 leads to $g_{12}^{(2)}(\tau) = g_{21}^{(2)}(\tau)$. 
Moreover, both correlations take their maximum values near zero delay and decrease as the delay time increases, demonstrating bunching between the two fluorescence photons within a short time window.
Figure~\ref{Fig3}(c) shows the delayed bundle correlation function $g_{\rm B}^{(2)}(\tau)$. Owing to the finite-level structure, $g_{\rm B}^{(2)}(0)=0$. The correlation remains below unity at short delays and gradually approaches unity at longer delays, indicating antibunching between successive photon bundles.
Together, the results in Figs.~\ref{Fig3}(a)--\ref{Fig3}(c) demonstrate that the two fluorescence photons are bunched within each individual bundle, whereas successive two-photon bundles are antibunched.

\begin{figure}[t]
  \centering
	\includegraphics[scale=1]{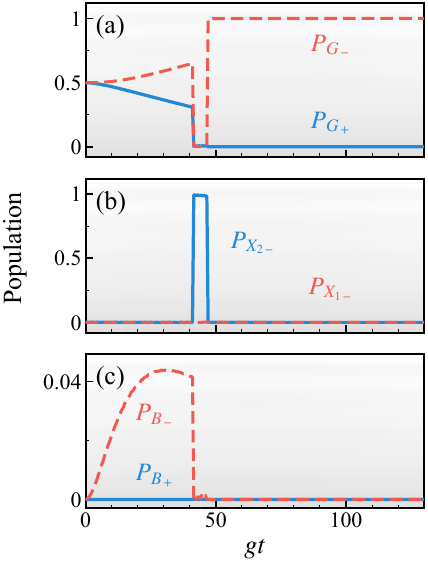}
\caption{ Quantum trajectory for the  fluorescence photon bundle emission. The initial state of the system is $|g_1g_2,g_3\rangle$, and the parameters are $\Delta=\Omega=30g$, $\gamma=0.1g$ and $\gamma_3=0.1\gamma$.
\label{Fig4}}
\end{figure}

To visualize the complete fluorescence photon bundle emission process, we employ Monte Carlo simulations to record a representative quantum trajectory of the system. 
Starting from the initial state $|g_1 g_2, g_3\rangle$, the system builds up a small population of $\sim 0.04$ in $|B_-\rangle = |e_1 e_2, -\rangle$ under the resonance condition, as shown in Fig.~\ref{Fig4}(c).
Once the system reaches this state, two sequential quantum jumps from the target emitters can occur. 
The order of the two emissions is probabilistic. In the representative trajectory of Fig.~\ref{Fig4}(b), emitter~1 emits the first photon and the system collapses to $|X_{2-}\rangle=|g_1e_2,-\rangle$ with near-unit probability. 
Emitter~2 then emits the second photon within a short delay and brings the system to $|G_-\rangle=|g_1g_2,-\rangle$, ensuring strong temporal correlation between the two photons. 
Finally, emitter~3 radiates an auxiliary photon and  resets the system to initial state $|g_1 g_2, g_3\rangle$. 
Due to its much lower decay rate, this emission occurs after a considerably longer waiting time, naturally separating successive photon bundles.

Taken together, the correlation-function analysis and quantum-trajectory simulation confirm the generation of fluorescence two-photon bundles. Since the finite-level structure intrinsically excludes higher-excitation manifolds, the dominant remaining imperfection arises from the residual population of the off-resonant single-excitation states, whose radiative decay produces undesired single photons. In the next section, we analyze this single-excitation leakage and show how it can be suppressed to improve the photon-bundle purity.

%==============================================================================================================
\section{Suppression of single-excitation leakage}
\label{sec4}

The above results demonstrate the generation of fluorescence two-photon bundles through the correlated excitation and subsequent emission of the two target emitters. 
We now focus on the main imperfection that limits the bundle purity, namely the residual population of the off-resonant single-excitation states $|X_{1\pm}\rangle$ and $|X_{2\pm}\rangle$. 
These states compete with the target two-excitation state $|B_-\rangle$ during the dissipative dynamics and may give rise to undesired single-photon emission.

To quantify the influence of the off-resonant single-excitation states on the desired emission channel, we define the total residual population in the single-excitation manifold as $P_{\rm single}=P_{X_{1-}}+P_{X_{1+}}+P_{X_{2-}}+P_{X_{2+}},$
where $P_{X_{j\pm}}=|C_{X_{j\pm}}|^2$, and introduce the instantaneous rate-weighted branching ratio
\begin{equation}
R=\frac{(\gamma_1+\gamma_2)P_{B_-}}{(\gamma_1+\gamma_2)P_{B_-}+\gamma_1(P_{X_{1-}}+P_{X_{1+}})
+\gamma_2(P_{X_{2-}}+P_{X_{2+}})},
\label{eq:R_def}
\end{equation}
with $P_{B_-}=|C_{B_-}|^2$. The intermediate-state amplitudes are obtained perturbatively in Appendix~\ref{app:residual_population}. 
For equal target-emitter decay rates, $\gamma_1=\gamma_2=\gamma$, Eq.~\eqref{eq:R_def} reduces to
\begin{equation}
R=\frac{2P_{B_-}}{2P_{B_-}+P_{\rm single}}.
\label{Eq:R}
\end{equation}
Before a target photon is emitted, the system can decay either from the desired two-excitation state $|B_-\rangle$ or from the weakly populated off-resonant single-excitation states. The former initiates the desired two-photon cascade, whereas the latter gives rise to undesired single-photon emission. Importantly, once the first photon is emitted from $|B_-\rangle$, the single-excitation states that were unwanted leakage channels before the emission become the desired intermediate states of the cascade, from which the second photon is emitted to complete the bundle. Therefore, the branching ratio $R$ directly quantifies the competition between the bundle-emission and single-photon channels and provides a natural measure of the photon-bundle purity.

To obtain a transparent analytical expression for the single-excitation leakage, we neglect dissipation and adopt the symmetric resonant parameters $\Delta_1=\Delta_2=\Delta=\Omega$, $g_1=g_2=g$, and $\Delta_3=0$. Dissipation will be included later in the full dissipative dynamics. Under these conditions, the driving strength $\Omega$ sets both the detuning and the dressed-state splitting, since $2\Delta=\Omega_R=2\Omega$. Thus, $\Omega$ serves as the key control parameter for suppressing the residual single-excitation population by increasing the energy separation of the off-resonant intermediate states. More general parameter choices may change the quantitative details but do not alter this physical picture. The four intermediate-state amplitudes then reduce to
\begin{eqnarray}
C_{X_{j-}} &\simeq& \frac{g}{2\Omega}\left(C_{G_+}-C_{B_-}\right), \nonumber\\
C_{X_{j+}} &\simeq& -\frac{g}{2\Omega}\left(C_{G_+}+C_{B_-}\right), \qquad j=1,2.
\end{eqnarray}
Consequently, the total single-excitation population becomes
$P_{\rm single}\simeq\frac{g^2}{\Omega^2}\left(|C_{G_+}|^2+|C_{B_-}|^2\right)$. Substituting this result into the equal-decay form of Eq.~\eqref{Eq:R}, for $P_{B_-}\neq0$, we obtain
\begin{eqnarray}
R=\frac{1}{1+\frac{g^2}{2\Omega^2}\left(1+{|C_{G_+}|^2}/{|C_{B_-}|^2}\right)}.
\label{eq:R_compact}
\end{eqnarray}

To obtain the explicit time dependence of $R$, we use the coherent dynamics governed by the effective two-state Hamiltonian. Although no actual emission occurs in this coherent evolution, the instantaneous populations determine the relative weights of the desired bundle-emission and single-excitation leakage channels prior to a target-emitter quantum jump. For the initial state $|G_+\rangle$, substituting the effective two-state dynamics into Eq.~\eqref{eq:R_compact} gives
\begin{equation}
R=\frac{1}{1+\frac{g^2}{2\Omega^2}\left[1+\cot^2\!\left(\frac{g^2t}{\Omega}\right)\right]}.
\label{eq:R_analytic}
\end{equation}
The resulting analytical expression is then benchmarked against the coherent dynamics generated by the full Hamiltonian.

To make the parameter dependence of $R$ more transparent, we introduce the dimensionless effective time $\tau=g^2t/\Omega$. Thus, the above equation can be rewritten as $R={1}/(1+\frac{g^2}{2\Omega^2}\csc^2\tau).$
This expression separates the effective evolution stage, characterized by $\tau$, from the leakage factor $g^2/(2\Omega^2)$. Thus, different driving strengths should be compared at the same effective time $\tau$. At fixed $\tau$, $R$ increases monotonically with $\Omega$, reflecting the suppression of the off-resonant single-excitation leakage.

At the effective transfer maximum of the target two-excitation state $|B_-\rangle$, corresponding to $\tau=\pi/2$, we obtain
\begin{equation}
R_{\max}=\frac{2\Omega^2}{2\Omega^2+g^2}\simeq1-\frac{g^2}{2\Omega^2},\qquad\Omega\gg g.
\label{eq:Rmax}
\end{equation}
This equation clearly shows the different influences of $\Omega$ and $g$ on the bundle purity. For a fixed coupling strength $g$, increasing the driving strength $\Omega$ suppresses the residual single-excitation leakage and monotonically increases $R_{\max}$ toward unity. When $\Omega$ is already large, further increasing the driving strength leads to only a marginal improvement in the bundle purity, consistent with the scaling $1-R_{\max}\simeq g^2/(2\Omega^2)$.
On the other hand, the dependence of $R_{\max}$ on $g$ enters through the small parameter $g^2/(2\Omega^2)$. Therefore, in the strong-driving regime, the direct effect of $g$ on $R_{\max}$ remains weak. Nevertheless, $g$ still affects the overall excitation and emission dynamics through the effective coupling $|g_{\rm eff}^{1}|\sim g^2/\Omega$.

\begin{figure}[t]
  \centering
  \includegraphics[scale=0.8]{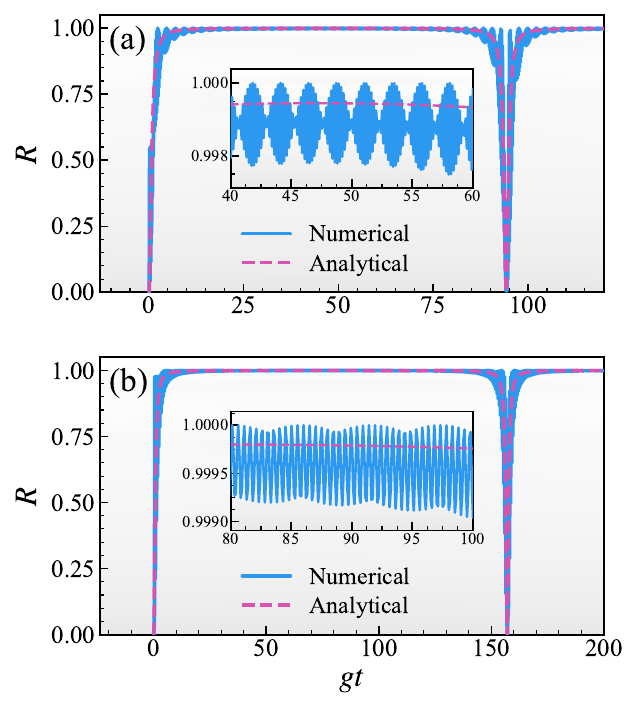}
\caption{
Time evolution of the parameter $R$ in the absence of system dissipation. The system is initially prepared in $|g_1g_2,+\rangle$. The numerical results are obtained from the full Hamiltonian in Eq.~\eqref{Eq1}, while the analytical results are calculated from Eq.~\eqref{eq:R_analytic}. The insets provide enlarged views of the corresponding regions. The driving strengths and detunings are (a) $\Omega=\Delta=30g$ and (b) $\Omega=\Delta=50g$.
\label{Fig5}
}
\end{figure}

\begin{figure}[t]
  \centering
	\includegraphics[scale=0.75]{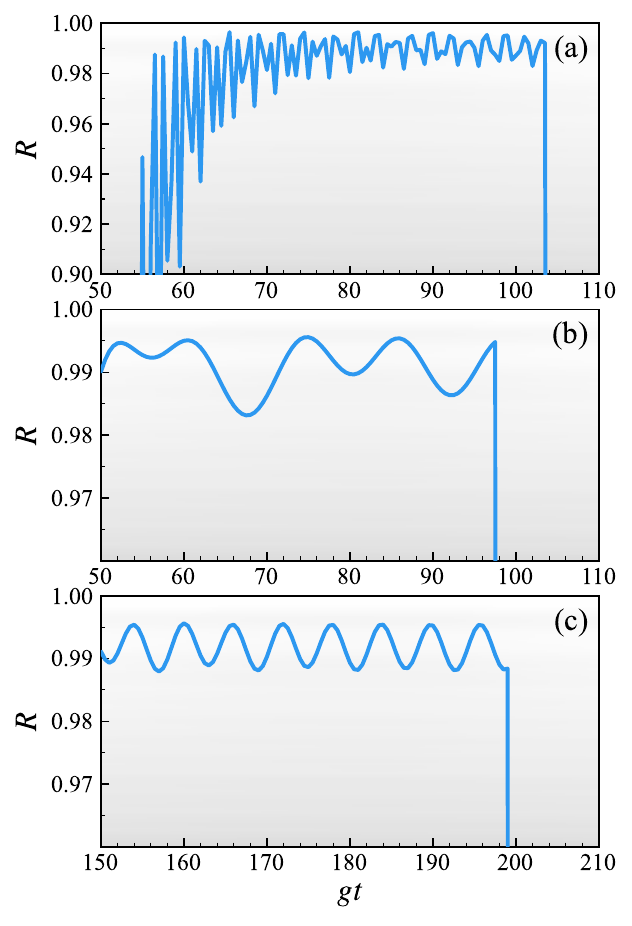}
\caption{
Time evolution of the parameter $R$ in the presence of system dissipation up to the occurrence of a target-emitter quantum-jump. 
Panels (a)--(c) correspond to $\Omega=30g$, $50g$, and $100g$, and other parameters are $\gamma=0.1g$ and $\gamma_3=0.1\gamma$.
\label{Fig6}
}
\end{figure}

Figure~\ref{Fig5} verifies the above analytical conclusions by comparing Eq.~\eqref{eq:R_analytic} with the numerical result obtained from the full Hamiltonian in Eq.~\eqref{Eq1}. 
For both driving strengths, $\Omega=30g$ and $\Omega=50g$, the analytical curves capture the overall behavior of the numerical results, confirming the validity of the perturbative expression for $R(t)$ in the absence of dissipation. 
The small-amplitude rapid oscillations in the numerical curves are absent from Eq.~\eqref{eq:R_analytic}, because the off-resonant single-excitation amplitudes are treated adiabatically and expressed in terms of the slowly varying amplitudes of $|G_+\rangle$ and $|B_-\rangle$. Their independent fast coherent dynamics and higher-order corrections are therefore not captured by the analytical expression.
The rapid drops of $R$ occur when the population of the target state $|B_-\rangle$ approaches zero, where the term $|C_{G_+}|^2/|C_{B_-}|^2$ in Eq.~\eqref{eq:R_compact} becomes very large. 
Apart from these points, $R$ remains close to unity, indicating that the residual population in the single-excitation manifold is strongly suppressed. 
Moreover, the insets of Figs.~\ref{Fig5}(a) and \ref{Fig5}(b) show a slight increase of the maximum $R$ as $\Omega$ increases from $30g$ to $50g$, consistent with Eq.~\eqref{eq:Rmax}. The small difference between the two cases indicates that $R_{\max}$ has already approached unity in the strong-driving regime.
\begin{figure}[t]
  \centering
	\includegraphics[scale=0.83]{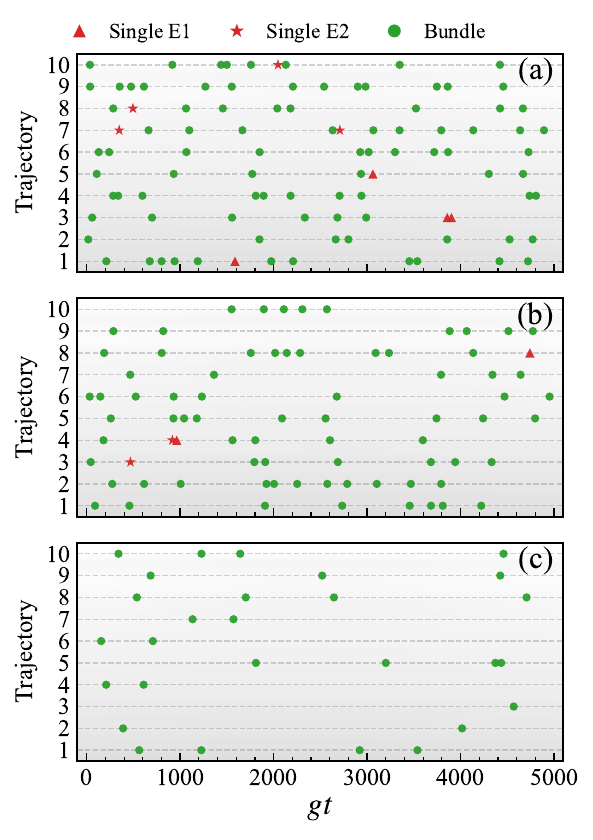}
\caption{ Quantum trajectories for different driving strengths. 
Green circles denote the target two-photon bundle emission events, while red triangles and red stars denote undesired single-photon emission events from emitter~1 and emitter~2, respectively. 
The driving strengths are (a) $\Omega=30g$, (b) $\Omega=50g$, and (c) $\Omega=100g$, and other parameters are the same as those in Fig.~\ref{Fig6}.
\label{Fig7}
}
\end{figure} 

\begin{figure*}[t]
  \centering
  \includegraphics[width=\textwidth]{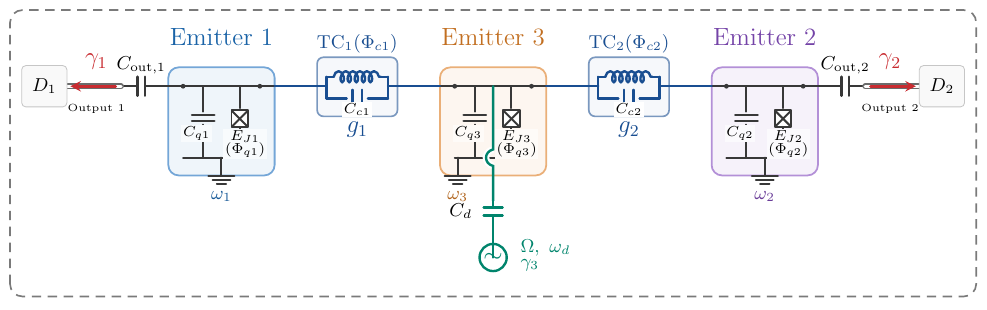}
\caption{Schematic of a possible superconducting-circuit implementation based on three flux-tunable transmon-type artificial atoms. Emitter~3 is coherently driven through a capacitively coupled microwave line and is coupled to emitters~1 and 2 through two independently controlled tunable couplers $TC_1(\Phi_{c1})$ and $TC_2(\Phi_{c2})$, which generate the effective exchange couplings $g_1$ and $g_2$, respectively. The fluorescence photons emitted by the two target emitters are coupled into independent output transmission lines through $C_{\rm out,1}$ and $C_{\rm out,2}$, with radiative decay rates $\gamma_1$ and $\gamma_2$, and are monitored by $D_1$ and $D_2$. The drive line also provides a weak radiative decay channel for emitter~3 with rate $\gamma_3$.}
\label{Fig8}
\end{figure*}

The above analysis is derived in the absence of dissipation. However, it captures the perturbative origin of single-excitation leakage and the main parameter dependence of $R$. Importantly, the parameter dependence identified in the nondissipative case remains valid in the presence of dissipation, providing a reliable guide for interpreting the bundle purity in the actual dynamics.
We next examine the behavior of $R$ in the presence of dissipation. 
Figure~\ref{Fig6} shows the time evolution of $R$ before the occurrence of a target-emitter quantum jump for different driving strengths. As $\Omega$ increases from $30g$ to $100g$, $R$ remains progressively closer to unity, confirming that the residual population of the off-resonant single-excitation states is increasingly suppressed. The oscillations of $R$ also become weaker at larger $\Omega$. These dissipative quantum-trajectory results are consistent with the analytical prediction that increasing the driving strength suppresses the single-excitation leakage and improves the photon-bundle purity.

To further visualize how the driving strength affects the emission events, Fig.~\ref{Fig7} shows representative quantum trajectories for different values of $\Omega$. 
The green circles denote the target two-photon bundle emission events, while the red triangles and red stars denote undesired single-photon emissions from emitter~1 and emitter~2, respectively. 
For $\Omega=30g$, several single-photon events appear in the trajectories, indicating that the residual population of the single-excitation manifold can still induce non-bundle emission. 
As the driving strength is increased to $\Omega=50g$ and $100g$, these undesired single-photon events are gradually suppressed, demonstrating the reduction of single-excitation leakage.
Meanwhile, Fig.~\ref{Fig7} also shows that the number of target two-photon bundles decreases with increasing $\Omega$. 
This behavior is consistent with the scaling $|g_{\rm eff}^{1}|\sim g^2/\Omega$: a stronger driving field suppresses the off-resonant intermediate states, but also weakens the effective excitation of $|B_-\rangle$. 
Therefore, increasing $\Omega$ improves the bundle purity at the cost of reducing the bundle emission number, revealing a trade-off between purity and emission rate. The present results, while derived for two-level emitters, are expected to qualitatively capture the role of intermediate-state leakage in bundle emission for bosonic systems.

Finally, we briefly discuss a possible implementation of the present scheme in superconducting quantum circuits.
A representative circuit based on three flux-tunable transmon qubits is shown in Fig.~\ref{Fig8}, while fluxonium qubits may be employed as an alternative platform with larger anharmonicity.
The three two-level emitters can be realized by superconducting artificial atoms operated in their lowest two levels.
Such platforms provide controllable transition frequencies, microwave driving fields, and engineered radiative decay channels through coupling to transmission lines or resonators.
In the representative circuit shown in Fig.~\ref{Fig8}, emitter~3 is coupled to the two target emitters through two independently controlled tunable couplers, while the direct coupling between emitters~1 and~2 is minimized by circuit design.
A closely related three-qubit coupling configuration has been experimentally realized using three frequency-tunable transmon qubits, in which one qubit is effectively coupled to each of the other two through tunable couplers \cite{Li2024ThreeQubit}.
Alternatively, the required coherent interactions may be implemented through direct capacitive or inductive coupling.
Tunable qubit--qubit couplings have been experimentally demonstrated in superconducting circuits \cite{Niskanen2007,Bialczak2011}, and recent fluxonium devices have reached tunable coupling strengths of several tens of megahertz \cite{Zhang2024}.
The fluorescence photons from emitters~1 and~2 are directed into two independent output transmission lines, allowing their emission events and cross correlations to be measured separately.
Resonance fluorescence from superconducting artificial atoms has been observed experimentally, with microwave photons emitted into transmission lines characterized through spectral, dynamical, and correlation measurements \cite{Toyli2016,Ruan2024}.
As a representative parameter set, one may take $g_1/2\pi=g_2/2\pi\simeq10~{\rm MHz}$ and engineered radiative decay rates $\gamma_1/2\pi=\gamma_2/2\pi\simeq1~{\rm MHz}$, with $\gamma_3\simeq0.1\gamma_{1,2}$, corresponding to $\gamma=0.1g$ in our simulations.
The strong-driving regime $\Omega/g=30$--$100$ then corresponds to $\Omega/2\pi\simeq300$--$1000~{\rm MHz}$, which can be approached using strongly driven artificial atoms with sufficiently large anharmonicity.
For comparison, recent tantalum-based transmon devices have achieved lifetimes and coherence times exceeding $0.3~{\rm ms}$, corresponding to intrinsic decoherence rates in the kilohertz range, well below the engineered megahertz-scale radiative decay rates considered here \cite{Place2021}.
These parameters indicate that the required regime of strong coherent driving, controllable qubit--qubit coupling, and engineered radiative decay with $\gamma/g\sim0.1$ is within reach of current superconducting-circuit technology.

\section{Conclusion}
\label{sec5}

In summary, we have proposed a scheme for generating fluorescence two-photon bundles using three coupled two-level emitters, where two target emitters emit strongly correlated fluorescence photons and a strongly driven auxiliary emitter mediates their correlated excitation. Unlike bosonic-mode-based schemes, the finite excitation space of the two target emitters intrinsically excludes higher-excitation manifolds and thereby eliminates contamination from undesired higher-excitation states. We further identified the residual population of the off-resonant single-excitation states as the dominant source of single-photon impurity and showed that this leakage can be effectively suppressed by increasing the driving strength, thereby improving the photon-bundle purity. At the same time, the reduced effective two-emitter coupling at strong driving lowers the bundle emission rate, leading to a trade-off between bundle purity and emission rate. 
This scheme provides a practical platform for generating high-purity, controllable fluorescence photon bundles, with potential applications in quantum information processing.

\section*{Acknowledgments}
This work is supported by the National Natural Science Foundation of China (Grants Nos. 12274164, 61875067, 12375011 and 12304392) and the Fundamental Research Funds for the Central Universities (Grant No. XJ2026002801).

%==============================================================================================================

\begin{appendix}
\section{Derivation of the effective Hamiltonian and residual intermediate-state populations}
\label{secA}
In this Appendix, we derive the effective Hamiltonian for the resonant two-excitation transition and evaluate the residual populations of the off-resonant intermediate states.

\subsection{Effective Hamiltonian}
\label{app:effective_hamiltonian}
In the dressed-state representation, the system Hamiltonian in Eq.~\eqref{Eq1} can be written as $H=H_0+V$, with
\begin{align}    
H_0 &= \sum_{j=1,2}\Delta_j \sigma_j^\dagger\sigma_j
      + \sum_{\eta=\pm}\lambda_\eta\sigma_{\eta\eta},
      \\
V   &= \sum_{j=1,2}g_j\sigma_j\Bigl(
        sc\sigma_{++}
        -s^2\sigma_{-+}
        +c^2\sigma_{+-}
        -sc\sigma_{--}
      \Bigr)
      +\textrm{H.c.}.
      \nonumber
\end{align}
where $s=\sin\theta$ and $c=\cos\theta$.
Since the system consists of three two-level emitters, the Hilbert space is eight dimensional. 
In the dressed-state representation of the driven emitter 3, we choose the following basis states and their corresponding unperturbed eigenenergies:

\begin{table} [http]
\caption{Basis states, notations, and the corresponding unperturbed eigenenergies in the dressed-state representation.}
\label{tab:basis_energy}
\begin{ruledtabular}
\begin{tabular}{ccc}
Basis state & Notation & Unperturbed energy \\
$|g_1 g_2,-\rangle$ & $|G_-\rangle$ & $\lambda_-$ \\
$|g_1 g_2,+\rangle$ & $|G_+\rangle$ & $\lambda_+$ \\
$|e_1 g_2,-\rangle$ & $|X_{1-}\rangle$ & $\Delta_1+\lambda_-$ \\
$|e_1 g_2,+\rangle$ & $|X_{1+}\rangle$ & $\Delta_1+\lambda_+$ \\
$|g_1 e_2,-\rangle$ & $|X_{2-}\rangle$ & $\Delta_2+\lambda_-$ \\
$|g_1 e_2,+\rangle$ & $|X_{2+}\rangle$ & $\Delta_2+\lambda_+$ \\
$|e_1 e_2,-\rangle$ & $|B_-\rangle$ & $\Delta_1+\Delta_2+\lambda_-$ \\
$|e_1 e_2,+\rangle$ & $|B_+\rangle$ & $\Delta_1+\Delta_2+\lambda_+$
\end{tabular}
\end{ruledtabular}
\end{table}

Using the ordered basis listed in Table~\ref{tab:basis_energy}, the Hamiltonian H can be represented in matrix form as
{\setlength{\arraycolsep}{7pt}%
\renewcommand{\arraystretch}{1.25}%
\begin{widetext}
\begin{equation}
\begin{pmatrix}
E_{G_-} & 0 & -g_1cs & -g_1s^2 & -g_2cs & -g_2s^2 & 0 & 0 \\
0 & E_{G_+} & g_1c^2 & g_1cs & g_2c^2 & g_2cs & 0 & 0 \\
-g_1cs & g_1c^2 & E_{X_{1-}} & 0 & 0 & 0 & -g_2cs & -g_2s^2 \\
-g_1s^2 & g_1cs & 0 & E_{X_{1+}} & 0 & 0 & g_2c^2 & g_2cs \\
-g_2cs & g_2c^2 & 0 & 0 & E_{X_{2-}} & 0 & -g_1cs & -g_1s^2 \\
-g_2s^2 & g_2cs & 0 & 0 & 0 & E_{X_{2+}} & g_1c^2 & g_1cs \\
0 & 0 & -g_2cs & g_2c^2 & -g_1cs & g_1c^2 & E_{B_-} & 0 \\
0 & 0 & -g_2s^2 & g_2cs & -g_1s^2 & g_1cs & 0 & E_{B_+}
\end{pmatrix}
\label{eq:H_matrix}
\end{equation}
\end{widetext}
}
The diagonal elements are the unperturbed eigenenergies given in Table~\ref{tab:basis_energy}.

As shown in Eq.~\eqref{eq:H_matrix}, there is no direct coupling between the collective ground states and the collective two-excitation states of emitters 1 and 2. 
The effective coupling is therefore generated by virtual transitions through the four single-excitation states $|X_{1\pm}\rangle$ and $|X_{2\pm}\rangle$. 
This leads to two resonant transition channels, $|G_+\rangle\leftrightarrow |B_-\rangle$ for $\Delta_1+\Delta_2=\Omega_R$ and $|G_-\rangle\leftrightarrow |B_+\rangle$ for $\Delta_1+\Delta_2=-\Omega_R$.

We first consider the channel $|G_+\rangle\leftrightarrow |B_-\rangle$. 
The resonant subspace is spanned by $\{|G_+\rangle,|B_-\rangle\}$, whereas the four states
$\mathcal{Q}=\{|X_{1-}\rangle,|X_{1+}\rangle,|X_{2-}\rangle,|X_{2+}\rangle\}$
serve as off-resonant intermediate states. 
By second-order perturbation theory, the effective coupling is
\begin{equation}
g^1_{\rm eff}=\sum_{\mu\in\mathcal{Q}}\frac{\langle G_+|V|\mu\rangle\langle \mu|V|B_-\rangle}{E_{G_+}-E_\mu}.
\label{Jeff_plus_general}
\end{equation}
Using the matrix elements in Eq.~\eqref{eq:H_matrix}, one obtains
\begin{equation}
g_{\rm eff}^{1}=-g_1g_2sc^3\left[\frac{1}{\Omega_R-\Delta_1}+\frac{1}{\Delta_1}+\frac{1}{\Omega_R-\Delta_2}+\frac{1}{\Delta_2}\right].
\label{eq:geff1_general}
\end{equation}

Thus, the effective Hamiltonian for the first resonant channel is
\begin{equation}
H_{\rm eff}^{1}=g_{\rm eff}^{1}\left(\sigma_1^\dagger \sigma_2^\dagger \sigma_{-+}+\sigma_1 \sigma_2 \sigma_{+-}\right),
\label{eq:Heff1_operator}
\end{equation}

Similarly, for the channel $|G_-\rangle\leftrightarrow |B_+\rangle$, the effective Hamiltonian is
\begin{equation}
H_{\rm eff}^{2}=g_{\rm eff}^{2}
\left(\sigma_1^\dagger\sigma_2^\dagger\sigma_{+-}+\sigma_1\sigma_2\sigma_{-+}\right),
\label{eq:Heff2}
\end{equation}
with
\begin{equation}
g_{\rm eff}^{2}=-g_1g_2cs^3\left[\frac{1}{\Delta_1}-\frac{1}{\Delta_1+\Omega_R}
+\frac{1}{\Delta_2}-\frac{1}{\Delta_2+\Omega_R}\right].
\label{eq:geff2_general}
\end{equation}

\subsection{Residual populations of the intermediate states}
\label{app:residual_population}
Although the single-excitation states $|X_{1\pm}\rangle$ and $|X_{2\pm}\rangle$ are far off resonance and are perturbatively eliminated in the effective Hamiltonian, they can acquire small residual populations during the actual dynamics. 
These populations represent leakage into undesired single-excitation states, which can produce single-photon emission and reduce the purity of the target bundle emission. 
Below we calculate the residual populations of the four intermediate states.

To evaluate the residual population of the off-resonant intermediate states, we adiabatically eliminate their amplitudes. For the first resonant channel $|G_+\rangle\leftrightarrow|B_-\rangle$, the wave function can be expanded as
\begin{equation}
|\psi(t)\rangle=C_{G_+}(t)|G_+\rangle+C_{B_-}(t)|B_-\rangle+\sum_{\mu\in\mathcal{Q}} C_\mu(t)|\mu\rangle ,
\end{equation}
where $\mathcal{Q}=\{|X_{1-}\rangle,|X_{1+}\rangle,|X_{2-}\rangle,|X_{2+}\rangle\}$. 
Since the states in $\mathcal{Q}$ are far off resonance, their amplitudes can be adiabatically eliminated, giving
\begin{equation}
C_\mu(t)\simeq\frac{V_{\mu G_+}C_{G_+}(t)+V_{\mu B_-}C_{B_-}(t)}{E_{G_+}-E_\mu},
\label{eq:Cmu_general}
\end{equation}
where $V_{\mu G_+}=\langle \mu|V|G_+\rangle$ and $V_{\mu B_-}=\langle \mu|V|B_-\rangle$. 
The corresponding residual population is therefore
\begin{equation}
P_\mu(t)=|C_\mu(t)|^2 .
\label{eq:Pmu_general}
\end{equation}

For the first resonant channel $|G_+\rangle\leftrightarrow |B_-\rangle$, the four intermediate-state amplitudes are obtained as
\begin{eqnarray}
C_{X_{1-}}(t)&\simeq&\frac{g_1c^2 C_{G_+}(t)-g_2cs C_{B_-}(t)}{\Omega_R-\Delta_1},\nonumber\\
C_{X_{1+}}(t)&\simeq&-\frac{g_1cs C_{G_+}(t)+g_2c^2 C_{B_-}(t)}{\Delta_1}.\nonumber \\
C_{X_{2-}}(t)&\simeq&\frac{g_2c^2 C_{G_+}(t)-g_1cs C_{B_-}(t)}{\Omega_R-\Delta_2},\nonumber\\
C_{X_{2+}}(t)&\simeq&-\frac{g_2cs C_{G_+}(t)+g_1c^2 C_{B_-}(t)}{\Delta_2}.
\label{eq:intermediate_amplitudes}
\end{eqnarray}

\end{appendix}

%===============================================================================================================


\begin{thebibliography}{99}

\bibitem{Bennett2000}
C. H. Bennett and D. P. DiVincenzo,
Quantum information and computation,
Nature \textbf{404}, 247--255 (2000).

\bibitem{Couteau2023}
C. Couteau, S. Barz, T. Durt, T. Gerrits, J. Huwer, R. Prevedel, J. Rarity, A. Shields, and G. Weihs,
Applications of single photons to quantum communication and computing,
Nat. Rev. Phys. \textbf{5}, 326--338 (2023).

\bibitem{Kimble2008}
H. J. Kimble,
The quantum internet,
Nature \textbf{453}, 1023--1030 (2008).

\bibitem{Ladd2010}
T. D. Ladd, F. Jelezko, R. Laflamme, Y. Nakamura, C. Monroe, and J. L. O'Brien,
Quantum computers,
Nature \textbf{464}, 45--53 (2010).

\bibitem{Dowling2008}
J. P. Dowling,
Quantum optical metrology--the lowdown on high-NOON states,
Contemp. Phys. \textbf{49}, 125--143 (2008).

\bibitem{Joo2011}
J. Joo, W. J. Munro, and T. P. Spiller,
Quantum metrology with entangled coherent states,
Phys. Rev. Lett. \textbf{107}, 083601 (2011).

\bibitem{Qin2023}
J. Qin, Y.-H. Deng, H.-S. Zhong, L.-C. Peng, H. Su, Y.-H. Luo, J.-M. Xu, D. Wu, S.-Q. Gong, H.-L. Liu, H. Wang, M.-C. Chen, L. Li, N.-L. Liu, C.-Y. Lu, and J.-W. Pan,
Unconditional and robust quantum metrological advantage beyond NOON states,
Phys. Rev. Lett. \textbf{130}, 070801 (2023).

\bibitem{Giovannetti2011}
V. Giovannetti, S. Lloyd, and L. Maccone,
Advances in quantum metrology,
Nat. Photonics \textbf{5}, 222--229 (2011).

\bibitem{Paris2009}
M. G. A. Paris,
Quantum estimation for quantum technology,
Int. J. Quantum Inf. \textbf{7}, 125--137 (2009).

\bibitem{MeyerScott2022}
E. Meyer-Scott, N. Prasannan, I. Dhand, C. Eigner, V. Quiring, S. Barkhofen, B. Brecht, M. B. Plenio, and C. Silberhorn,
Scalable generation of multiphoton entangled states by active feed-forward and multiplexing,
Phys. Rev. Lett. \textbf{129}, 150501 (2022).

\bibitem{Gisin2007}
N. Gisin and R. Thew,
Quantum communication,
Nat. Photonics \textbf{1}, 165--171 (2007).

\bibitem{Sangouard2011}
N. Sangouard, C. Simon, H. de Riedmatten, and N. Gisin,
Quantum repeaters based on atomic ensembles and linear optics,
Rev. Mod. Phys. \textbf{83}, 33--80 (2011).

\bibitem{Llewellyn2020}
D. Llewellyn, Y. Ding, I. I. Faruque, S. Paesani, D. Bacco, R. Santagati, Y.-J. Qian, Y. Li, Y.-F. Xiao, M. Huber, M. Malik, G. F. Sinclair, X. Zhou, K. Rottwitt, J. L. O'Brien, J. G. Rarity, Q. Gong, L. K. Oxenløwe, J. Wang, and M. G. Thompson,
Chip-to-chip quantum teleportation and multi-photon entanglement in silicon,
Nat. Phys. \textbf{16}, 148--153 (2020).

\bibitem{Horton2013}
N. G. Horton, K. Wang, D. Kobat, C. G. Clark, F. W. Wise, C. B. Schaffer, and C. Xu,
In vivo three-photon microscopy of subcortical structures within an intact mouse brain,
Nat. Photonics \textbf{7}, 205--209 (2013).

\bibitem{Li2023}
S. Li, R. Chang, L. Zhao, R. Xing, J. C. M. van Hest, and X. Yan,
Two-photon nanoprobes based on bioorganic nanoarchitectonics with a photo-oxidation enhanced emission mechanism,
Nat. Commun. \textbf{14}, 5227 (2023).

\bibitem{munoz2014emitters}
C. S. Muñoz, E. Del Valle, A. González-Tudela, K. Müller, S. Lichtmannecker, M. Kaniber, C. Tejedor, J. J. Finley, and F. P. Laussy,
Emitters of $N$-photon bundles,
Nat. Photonics \textbf{8}, 550--555 (2014).

\bibitem{Munoz2018}
C. S. Muñoz, F. P. Laussy, E. Del Valle, C. Tejedor, and A. González-Tudela,
Filtering multiphoton emission from state-of-the-art cavity quantum electrodynamics,
Optica \textbf{5}, 14--26 (2018).

\bibitem{Gou2022}
C. Gou, X. Hu, and F. Wang,
Antibunched two-mode two-photon bundles via atomic coherence,
Phys. Rev. A \textbf{106}, 063718 (2022).

\bibitem{Gou2024NJP}
C. Gou, J. Xu, F. Wang, and X. Hu,
Antibunched $N$-photon bundles from dark states assisted by ac Stark shift,
New J. Phys. \textbf{26}, 073046 (2024).

\bibitem{Jiang2023}
S.-Y. Jiang, F. Zou, Y. Wang, J.-F. Huang, X.-W. Xu, and J.-Q. Liao,
Multiple-photon bundle emission in the $n$-photon Jaynes-Cummings model,
Opt. Express \textbf{31}, 15697--15711 (2023).

\bibitem{Tang2023}
J. Tang and Y.-G. Deng,
Strong single-photon to two-photon bundles emission in spin-1 Jaynes-Cummings model,
APL Photonics \textbf{8}, 076103 (2023).

\bibitem{Liu2023}
C. Liu, J.-F. Huang, and L. Tian,
Deterministic generation of multi-photon bundles in a quantum Rabi model,
Sci. China Phys. Mech. Astron. \textbf{66}, 220311 (2023).

\bibitem{Munoz2015}
C. S. Muñoz, F. P. Laussy, C. Tejedor, and E. Del Valle,
Enhanced two-photon emission from a dressed biexciton,
New J. Phys. \textbf{17}, 123021 (2015).

\bibitem{Cosacchi2022}
M. Cosacchi, A. Mielnik-Pyszczorski, T. Seidelmann, M. Cygorek, A. Vagov, D. E. Reiter, and V. M. Axt,
$N$-photon bundle statistics in different solid-state platforms,
Phys. Rev. B \textbf{106}, 115304 (2022).

\bibitem{GonzalezTudela2017}
A. González-Tudela, V. Paulisch, H. J. Kimble, and J. I. Cirac,
Efficient multiphoton generation in waveguide quantum electrodynamics,
Phys. Rev. Lett. \textbf{118}, 213601 (2017).

\bibitem{Xing2024}
F. Xing, Z. Liao, and X.-H. Wang,
Deterministic generation of arbitrary $n$-photon states in a waveguide-QED system,
Phys. Rev. A \textbf{109}, 013718 (2024).

\bibitem{Ma2021}
S.-L. Ma, X.-K. Li, Y.-L. Ren, J.-K. Xie, and F.-L. Li,
Antibunched $N$-photon bundles emitted by a Josephson photonic device,
Phys. Rev. Res. \textbf{3}, 043020 (2021).

\bibitem{Ma2022}
S.-L. Ma, J.-K. Xie, Y.-L. Ren, X.-K. Li, and F.-L. Li,
Photon-pair blockade in a Josephson-photonics circuit with two nondegenerate microwave resonators,
New J. Phys. \textbf{24}, 053001 (2022).

\bibitem{Zou2023}
F. Zou, Y. Li, and J.-Q. Liao,
Dynamical $N$-photon bundle emission,
New J. Phys. \textbf{25}, 043027 (2023).

\bibitem{Bin2020}
Q. Bin, X.-Y. Lü, F. P. Laussy, F. Nori, and Y. Wu,
$N$-phonon bundle emission via the Stokes process,
Phys. Rev. Lett. \textbf{124}, 053601 (2020).

\bibitem{Zou2022}
F. Zou, J.-Q. Liao, and Y. Li,
Dynamical emission of phonon pairs in optomechanical systems,
Phys. Rev. A \textbf{105}, 053507 (2022).

\bibitem{Deng2021}
Y. Deng, T. Shi, and S. Yi,
Motional $n$-phonon bundle states of a trapped atom with clock transitions,
Photon. Res. \textbf{9}, 1289--1299 (2021).

\bibitem{Yuan2023}
H. Y. Yuan, J. Xie, and R. A. Duine,
Magnon bundle in a strongly dissipative magnet,
Phys. Rev. Appl. \textbf{19}, 064070 (2023).

\bibitem{Liu2025OptLett}
J. Liu, S. Hu, W. Zhong, G. Cheng, and A. Chen,
Controllable antibunching of two-magnon bundle in a hybrid ferromagnet-superconductor system,
Opt. Lett. \textbf{50}, 682--685 (2025).

\bibitem{Gou2024PRR}
C. Gou, X. Hu, J. Xu, and F. Wang,
Hybrid magnon-photon bundle emission from a ferromagnetic-superconducting system,
Phys. Rev. Res. \textbf{6}, 023052 (2024).

\bibitem{Bin2024LPR}
Q. Bin, Q.-Y. Qiu, Y. Wu, and X.-Y. Lü,
Entangled photon-magnon bundle emission,
Laser Photonics Rev. \textbf{18}, 2300977 (2024).

\bibitem{Wang2025OL}
Z. Wang, W. Shi, D. Kong, H. Zhan, and F. Wang,
Photon--magnon bundle emission via enhanced photon--magnon--atom tripartite interaction,
Opt. Lett. \textbf{50}, 5386--5389 (2025).

\bibitem{Zhao2025PRA}
C. Zhao, W. Li, B. Xiong, J.-X. Peng, L. Zhou, and W.-J. Gong,
Heralded generation of entangled states based on $N$-bundle emission in a waveguide magnonics system,
Phys. Rev. A \textbf{112}, 013723 (2025).

\bibitem{Chang2016}
Y. Chang, A. González-Tudela, C. Sánchez Muñoz, C. Navarrete-Benlloch, and T. Shi,
Deterministic down-converter and continuous photon-pair source within the bad-cavity limit,
Phys. Rev. Lett. \textbf{117}, 203602 (2016).

\bibitem{Chen2025PRA}
Z. Chen, C. Gou, X. Hu, D. Kong, and F. Wang,
Enhanced bundle emission of squeezed photons using parametric amplification,
Phys. Rev. A \textbf{112}, 063703 (2025).

\bibitem{Bin2021}
Q. Bin, Y. Wu, and X.-Y. Lü,
Parity-symmetry-protected multiphoton bundle emission,
Phys. Rev. Lett. \textbf{127}, 073602 (2021).

\bibitem{Xiong2025}
B. Xiong, Q. Bin, S.-L. Chao, J.-B. Liu, and X.-Y. Lü,
Two-photon decay enhanced even photon bundle emission,
Phys. Rev. Res. \textbf{7}, 013238 (2025).

\bibitem{Chen2026Chiral}
Z. Chen, D. Kong, C. Gou, X. Hu, and F. Wang,
Chiral interaction enhanced magnon bundle emission,
Phys. Rev. A \textbf{114}, 013710 (2026).

\bibitem{Niskanen2007}
A. O. Niskanen, K. Harrabi, F. Yoshihara, Y. Nakamura, S. Lloyd, and J. S. Tsai,
Quantum coherent tunable coupling of superconducting qubits,
Science \textbf{316}, 723--726 (2007).

\bibitem{Bialczak2011}
R. C. Bialczak, M. Ansmann, M. Hofheinz, M. Lenander, E. Lucero, M. Neeley, A. D. O'Connell, D. Sank, H. Wang, M. Weides, J. Wenner, T. Yamamoto, A. N. Cleland, and J. M. Martinis,
Fast tunable coupler for superconducting qubits,
Phys. Rev. Lett. \textbf{106}, 060501 (2011).

\bibitem{Zhang2024}
H. Zhang, C. Ding, D. K. Weiss, Z. Huang, Y. Ma, C. Guinn, S. Sussman, S. P. Chitta, D. Chen, A. A. Houck, J. Koch, and D. I. Schuster,
Tunable inductive coupler for high-fidelity gates between fluxonium qubits,
PRX Quantum \textbf{5}, 020326 (2024).

\bibitem{Li2024ThreeQubit}
X.-L. Li, Z. Tao, K. Yi, K. Luo, L. Zhang, Y. Zhou, S. Liu, T. Yan, Y. Chen, and D. Yu,
Hardware-efficient and fast three-qubit gate in superconducting quantum circuits,
Front. Phys. \textbf{19}, 51205 (2024).

\bibitem{Toyli2016}
D. M. Toyli, A. W. Eddins, S. Boutin, S. Puri, D. Hover, V. Bolkhovsky, W. D. Oliver, A. Blais, and I. Siddiqi,
Resonance fluorescence from an artificial atom in squeezed vacuum,
Phys. Rev. X \textbf{6}, 031004 (2016).

\bibitem{Ruan2024}
X. Ruan, J.-H. Wang, D. He, P. Song, S. Li, Q. Zhao, L. M. Kuang, J.-S. Tsai, C.-L. Zou, J. Zhang, D. Zheng, O. V. Astafiev, Y.-x. Liu, and Z. Peng,
Dynamics and resonance fluorescence from a superconducting artificial atom doubly driven by quantized and classical fields,
Phys. Rev. Res. \textbf{6}, 033064 (2024).

\bibitem{Place2021}
A. P. M. Place, L. V. H. Rodgers, P. Mundada, B. M. Smitham, M. Fitzpatrick, Z. Leng, A. Premkumar, J. Bryon, S. Sussman, G. Cheng, T. Madhavan, H. K. Babla, B. Jaeck, A. Gyenis, N. Yao, R. J. Cava, N. P. de Leon, and A. A. Houck,
New material platform for superconducting transmon qubits with coherence times exceeding $0.3$ milliseconds,
Nat. Commun. \textbf{12}, 1779 (2021).

\end{thebibliography}
\end{document}